\documentclass[prx,twocolumn,floatfix,superscriptaddress]{revtex4-2}
\usepackage{dcolumn,amsmath}
\usepackage{graphicx}
\usepackage{braket}
\usepackage{bm}
\usepackage{amssymb}
\usepackage{hyperref}
\usepackage{multirow}
\usepackage{footnote}
\usepackage{ragged2e}
\usepackage{xcolor}
\usepackage{diagbox}
\usepackage{physics}
\usepackage{tabularx}

\newcommand{\twobytwo}[4]{ \begin{pmatrix}
   #1 & #2 \\
   #3 & #4 
  \end{pmatrix}} 
  
\newcommand{\onebytwo}[2]{ \begin{pmatrix}
   #1\\
   #2 
  \end{pmatrix}} 
\newcommand{\JQI}{Joint Quantum Institute, National Institute of Standards and Technology and University of Maryland, Gaithersburg, Maryland 20899, USA}
\newcommand{\SSD}{Sensor Science Division, National Institute of Standards and Technology, Gaithersburg, Maryland 20899, USA}
\newcommand{\NISTB}{National Institute of Standards and Technology (NIST), Boulder, CO 80305}

\begin{document}

\title{Quantum Blackbody Thermometry}

\author{Eric B. Norrgard}
\affiliation{\JQI}
\author{Stephen P. Eckel}
\affiliation{\SSD}
\author{Christopher L. Holloway}
\affiliation{\NISTB}
\author{Eric L. Shirley}
\affiliation{\SSD}

\date{\today}

\begin{abstract}
Blackbody radiation (BBR) sources are calculable radiation sources that are frequently used in radiometry, temperature dissemination, and remote sensing.   Despite their ubiquity, blackbody sources and radiometers have a plethora of systematics. We envision a new, primary route to measuring blackbody radiation using ensembles of polarizable quantum systems, such as Rydberg atoms and diatomic molecules. Quantum measurements with these exquisite electric field sensors 
could enable active feedback, improved design, and, ultimately, lower radiometric and thermal uncertainties of blackbody standards. A portable, calibration-free Rydberg-atom physics package could also complement a variety of classical radiation detector and thermometers. The successful merger of quantum and blackbody-based measurements provides a new, fundamental paradigm for blackbody physics.
\end{abstract}

\maketitle

\section{Introduction}\label{Sec1}

In 2019, the International System of Units (SI) 
was 
redefined in terms of a set of exact values of physical constants, replacing a system which included reference artifacts.  This new system allows anyone, in principle, to perform an identical measurement and arrive at the same value without prior coordination regarding instrumentation.  Along with the SI redefinition is a push toward a ``quantum SI,'' i.e. the ability to realize truly identical metrology by using identical quantum systems which are sensitive to the desired observable via immutable quantum behavior and other fundamental physical laws 
calculable from first principles.

Blackbodies are incoherent electromagnetic radiation sources that are ubiquitous in radiometry, temperature dissemination, and remote sensing.  Blackbody radiation is inherently  quantum in nature, as Planck famously hypothesized the quantum nature of light to explain the observed relation between blackbody spectral energy density and wavelength. Using Planck’s law, blackbodies establish a clear relationship between temperature and 
radiant 
power. 
This link allows for the calibration of RF noise, IR imagers, pyrometers, radiation thermometers, and other detectors. 

Despite their ubiquity, blackbody sources are susceptible to several systematic errors.  In particular, substantial offsets are often observed between measured radiance temperature and temperature measured via contact thermometers \cite{Carter2006},  with the uncertainty in the offset growing with time from calibration of the contact thermometer.  An important challenge is thus to develop a robust alternative to contact thermometers for probing blackbody references, especially for applications which preclude thermometer recalibration, such as remote-sensing satellites. Other systematic effects (e.g., emissivity, propagation loss, temperature gradients, geometric effects, etc.) may be important to blackbody performance as well.  

Furthermore, because  the detectors blackbodies calibrate are fundamentally classical, these calibrations typically involve undesirably 
long traceability chains.  For example, the International Temperature Scale of 1990 (ITS-90) \cite{PrestonThomas1990} defines temperatures above 1234.93\,K via radiation thermometry.  Here, a blackbody acts as a source of radiation which calibrates an optical detector.  However, standard BBR thermometers are entirely classical, typically 
involving an optical system with 
lenses, a monochromator or other spectral filters, integrating sphere, and detector (such as a photodiode).  Each of these classical elements must be carefully characterized in order to accurately measure radiative temperature.  For the temperature range 13.8033\,K to 1234.93\,K, \mbox{ITS-90} may determine temperature $T_{90}$ by any of 11 different interpolation functions for platinum resistance thermometers which are calibrated at specific defining fixed points; in this important temperature range, both the reference and detector are entirely classical.  Furthermore, these fixed point materials, as well as their containment vessels, have stringent purity requirements.

Atoms and molecules are 
immutable 
quantum systems whose interactions with electromagnetic radiation have been characterized with exquisite precision and, in many cases, are amenable to \textit{ab initio} calculation.  
Atomic transitions already are used explicitly to define the SI second, and implicitly to define all other SI base units apart from the mole.  
An atom- or molecule-based detector could thus provide internal temperature calibration to a blackbody reference to form a direct, fully quantum-SI realization of radiative temperature.

Blackbody radiation perturbs the internal quantum states of both atoms and molecules.  For instance, the BBR-induced Stark shift currently represents the largest, uncompensated systematic in optical clocks \cite{Beloy2014}. If used as a thermometer, optical clocks can currently measure temperatures to a fractional precision approaching $10^{-5}$ \cite{Beloy2014}. By choosing a quantum system with a larger polarizability, like a Rydberg atom  or a molecule, the Stark shift signal can be increased by a factor of $\sim10^3$ \cite{Figger1980,Ovsiannikov2011}.

\begin{figure*}
    \centering
    \includegraphics[width=\textwidth]{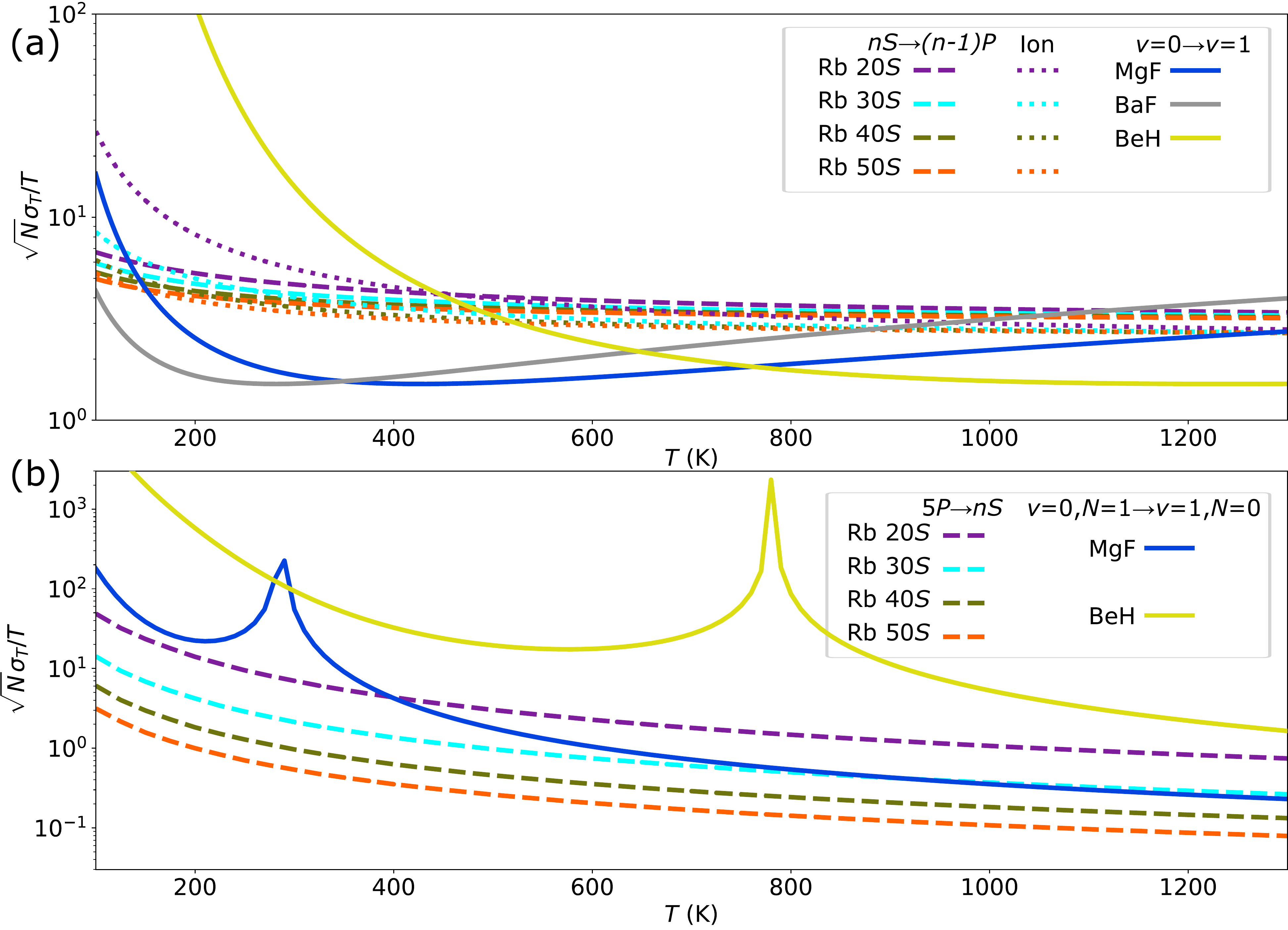}
    \caption{\label{fig:state_transfer_sensitivity}  Fractional temperature sensitivity as a function of temperature. (a)  State transfer measurements.  Solid lines show population thermalization between $v=0$ and $v=1$ for molecules. Dashed curves show the sensitivity of various $nS$ states of Rb, measuring the transfer to adjacent $P$ levels.  Dotted  curves show the sensitivity to BBR-induced ionization of Rb from  $nS$. (b) Lifetime-limited frequency shift measurements.  Solid lines show sensitivity of the $\ket{v=0,N=1}\rightarrow \ket{v=1,N=0}$ transition in molecules, dashed lines show the $5P\rightarrow nS$ transition in Rb.  }
\end{figure*}


Here, we consider several quantum measurement approaches using polar molecules and Rydberg atoms to determine the temperature $T$ of the surrounding blackbody radiation.  The main results of this work are estimates of the achievable  fractional temperature uncertainty $\sqrt{N}\sigma_T/T$ of each approach (in this work, $\sigma_x$ denote the standard uncertainty in variable $x$), which are summarized in Fig.~\ref{fig:state_transfer_sensitivity}. Assuming $N$ uncorrelated measurements, each measurement approach may potentially achieve $\sqrt{N}\sigma_T/T$ of order unity over a wide range of temperatures.    Such measurements would constitute direct measurements of thermodynamic temperature in a range currently practically realized under ITS-90 by interpolation between defining fixed points \cite{PrestonThomas1990} and correction for $T-T_{90}$ \cite{Fischer2011}. Section II  details the interaction of quantum systems with BBR 
that 
underpins the 
thermometric 
approaches.  
Section III estimates $\sqrt{N}\sigma_T/T$ for laser coolable diatomic molecules from BBR-induced state transfer and level shifts; Section IV applies the same considerations to Rydberg atoms.  Sections III and IV also provide experimental considerations relevant to each system.  Given the current state of the art, we estimate fractional temperature uncertainty $\sigma_T/T \approx 10^{-5}$  could be expected from molecule or Rydberg state transfer measurements or from Rydberg frequency shift measurements.  Frequency shift measurements of temperature in molecules are less competitive.


\section{Blackbody radiation and the two-level system}
Planck's law gives the spectral energy density $U(\omega,T)$ of an ideal blackbody at temperature $T$:
\begin{equation}
\begin{aligned}
    U(\omega, T) &= \frac{\hbar  \omega^3}{\pi^2 c^3} \frac{1}{e^{ \hbar\omega/k_{\rm{B}} T}-1}\\
    &=\frac{\hbar \omega ^3}{\pi^2 c^3}\left< n (\omega)\right>,
\end{aligned}
\end{equation}
where $\left< n (\omega)\right>$ is the mean photon number with frequency $\omega$.
Due to the narrow linewidths of the molecular vibrational transitions and atomic Rydberg transitions considered here, it is a good approximation to consider each pair of states $\ket{i}$ and $\ket{j}$ interacting with a single, resonant mode of the blackbody with frequency $\omega_{ij}$.  In this case, the spontaneous decay rate $\Gamma_{ij}$ is given by 
\begin{equation}\label{eq:spontaneous rate}
    \Gamma_{ij} = A_{ij} = \frac{\mu_{ij}^2 \omega_{ij}^3}{3 \epsilon_0 \hbar \pi c^3},
\end{equation}
and the stimulated rate $\Omega_{ij}$ is given by
\begin{equation}\label{eq:stimulated rate}
    \Omega_{ij}= B_{ij}U(\omega_{ij},T) = \frac{\mu_{ij}^2 \omega_{ij}^3}{3 \epsilon_0 \hbar \pi c^3} \langle n(\omega) \rangle,
\end{equation}
where $A_{ij}$ and $B_{ij}$ are the usual Einstein coefficients, and $\mu_{ij}^2 = \left|\Bra{i}d\ket{j}\right|^2$ is the 
dipole matrix element 
between states $\ket{i}$ and $\ket{j}$.  
%
%

Blackbody radiation 
also 
shifts the energy of a quantum state $\ket{i}$ by an amount \cite{Farley1981}
\begin{equation}
	\label{eq:full_freq_shift}
    \Delta E_i = \frac{1}{6\pi^2 \epsilon_0 c^3}\sum_j P\int_0^\infty d\omega  \frac{\omega^3\mu_{ij}^2}{e^\frac{\hbar\omega}{k_{\rm{B}} T}-1} \left(\frac{2\omega_{ij}}{\omega_{ij}^2-\omega^2}\right),
\end{equation}
with $P$ indicating the Cauchy principal value. Using the Farley-Wing function \cite{Farley1981},
\begin{equation}
    \mathcal{F}(y) = -2y P\int_0^\infty dx \frac{x^3}{(x^2-y^2)(e^x-1)},
\end{equation}
Eq.~\eqref{eq:full_freq_shift} becomes
\begin{equation}
    \label{eq:freq_shift}
	\Delta E_i = -\frac{1}{6\pi^2 \epsilon_0 c^3}\left(\frac{k_B T}{\hbar}\right)^3 \sum_j \mu_{ij}^2 \mathcal{F}\left(\frac{\hbar\omega_{ij}}{k_B T}\right).
\end{equation}
In the limit where $y\ll1$, the Farley-Wing function reduces to 
\begin{equation}
    \label{eq:farley_wing_approx}
    \mathcal{F}(y) \approx -\frac{\pi^2}{3}y.
\end{equation}
This limit is important in understanding the asymptotic forms that appear below. 


In the following sections, we apply the above equations  with generalizations to model the interactions of molecules and Rydberg atoms with BBR.

\section{Molecules}
Cold polar molecules are an emerging quantum technology with great promise for accurately probing BBR.      Vanhaecke and Dulieu have considered the state transfer and frequency shifting effects of blackbody radiation on polar molecules for use in precision measurements \cite{Vanhaecke2007a}.  Buhmann \textit{et al.} made similar consideration of the BBR state transfer rates for molecules in proximity to a surface.  Here, we interpret the state transfer and frequency shifts in terms of a BBR thermal measurement. Alyabyshev \textit{et al.} considered using polar molecules to measure applied electric fields \cite{Alyabyshev2012}.  Laser and frequency comb spectroscopy have demonstrated as low as 7\,mK precision in measuring the temperature of atmospheric CO$_2$ \cite{Hieta2011, Gianfrani2016, Hansel2017}.  The fractional accuracy $\sigma_T/T$ of such measurements in the field is limited to a few $10^{-4}$, however, by uncertainties in abundance and atmospheric pressure. 

Diatomic molecules are well suited to probing thermodynamic temperature as their vibrational transition frequencies are typically commensurate with the peak intensity of the blackbody spectrum around room temperature.  Several approaches have been developed over the last two decades to produce ultracold trapped molecules \cite{Carr2009}, including Stark and Zeeman decelerators \cite{Bochinski03,vandeMeerakker2005,Hoekstra2007, Narevicius2008}, optoelectric cooling \cite{Prehn2016}, magneto- and photoassociation of ultracold atoms \cite{ni2008,Danzl2010,Jones2006,Stellmer2012}, cryogenic buffer gas loading \cite{Weinstein1998,Stoll2008}, and direct laser cooling \cite{DiRosa2004,Stuhl2008,Barry2014,McCarron2018b}. 

Molecules which may be laser cooled are especially well-suited for BBR thermometry.  Unlike most molecules, this class of molecular species has optical cycling transitions, allowing for scattering of potentially millions of optical photons while remaining in one (or a few) quantum states.  In the following section, we show that optical cycling transitions may be used to thermalize the molecule population in only the lowest two vibrational levels with the surrounding BBR, enabling sensitive BBR thermometry.  Moreover, optical cycling can be used to perform efficient state readout.  Recent experiments \cite{McCarron2018b,Williams2018,Cheuk2018} with laser cooled molecules have demonstrated up to $N\sim10^4$ molecules at temperatures down to $T_{\rm{mol}}\sim10\,\mu$K.  These low temperatures could enable, e.g.\ a molecule fountain \cite{Tarbutt2013,Cheng2016} with molecule trajectories entering then exiting the interior of a reference blackbody cavity.

\subsection{State Transfer}
Previous considerations of BBR on trapped molecules have primarily focused on its effects on trap lifetime \cite{Hoekstra2007,Buhmann2008,Williams2018,chou2019}.  For molecules trapped in the ground vibrational state $v=0$, the relevant timescale for trap loss due to BBR effects is the inverse $v=0\rightarrow v^\prime= 1$  BBR-induced transition rate $1/\Omega_{01}$.  This is consistent with experimental observations: BBR has been reported to limit the trap lifetime for electrostatically trapped OH and OD \cite{Hoekstra2007} and magnetically trapped CaF \cite{Williams2018}.  Furthermore, BBR-induced rotational excitation is an important consideration in molecular ion clocks \cite{chou2019}.

\begin{figure}
    \centering
    \includegraphics[width=\columnwidth]{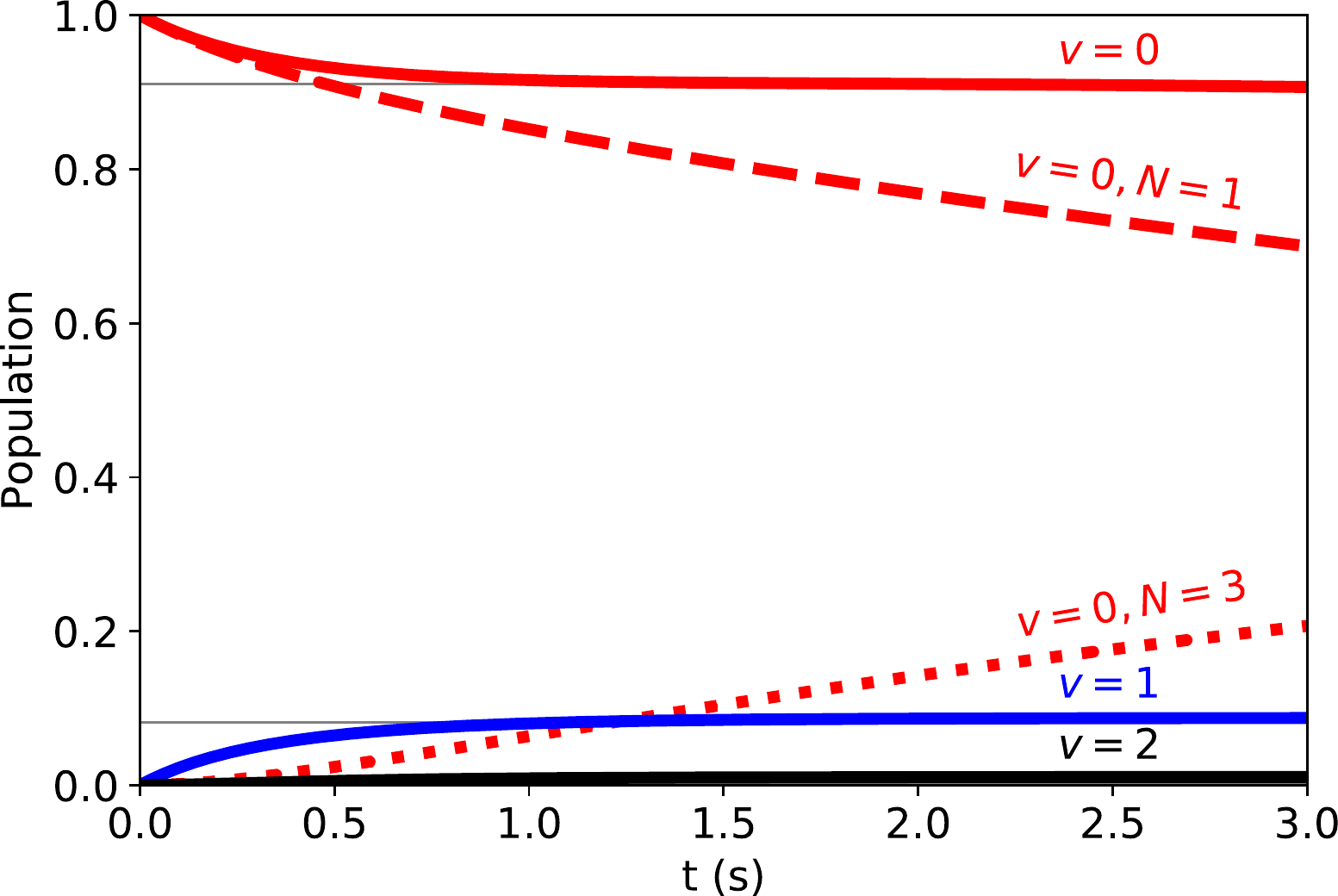}
    \caption{Rate equation simulation of free evolution in BBR field of SrF molecules ($v=0-2, N=0-5$, all hyperfine levels included).  The state is $\ket{v=0,N=1}$ at time $t\,=\,0$.  Solid red, blue, black lines denote the total $v=0,1,2$ populations, respectively.  Dashed (dotted) red lines show the $\ket{v=0,N=1}$ ($\ket{v=0,N=3}$) population.  Grey lines denote Maxwell-Boltzmann distribution of SrF vibrational levels at $T=300K$.}
    \label{fig:SrF simulation}
\end{figure}

However, determination of the BBR temperature via population transfer can be done on the faster timescale  $t_{\rm{therm}}= 1/(\Gamma_{01}+2\Omega_{01})$. This is illustrated in Fig.\,\ref{fig:SrF simulation} where we model the SrF system in a 300\,K BBR field.
We perform rate equation calculations for the populations of a Hund's case b \cite{Brown2003} ground state with total nuclear spin $\boldsymbol{I}=1/2$. 
Included in the simulation are all hyperfine and Zeeman sublevels in vibrational states $v=0-4$ and rotational states $N=0-5$.
with population initially in the $\ket{v=0, N^P=1^-,F=2, m_F=2}$ state ($P=\pm$ is the state's spatial parity, 
$F$ is the total angular momentum of the system and $m_F$ is its projection along the quantization axis).
In this case, $1/\Gamma_{01}\approx 0.36$\,s, $1/\Omega_{01} \approx 3.7$\,s, and $t_{\rm{therm}} \approx 0.33$\,s.
The total population in each vibrational manifold (solid lines) approaches those predicted by a classical thermodynamic equilibrium (grey lines) after roughly $t_{\rm{therm}}$, while the $\ket{v=0,N=1}$ population (red dashed line) continues to be excited to higher rovibrational states (red dotted line) for time $t~>~t_{\rm{therm}}$.

While the vibrational state populations quickly thermalize in time of order $t_{\rm{therm}}$, within  each vibrational manifold the state population will continue to redistribute over a ladder of rotational states $N = 0,1,2,\dots$.  The continuous excitation to higher rotational states within a vibrational manifold presents a challenge to determining the temperature of the BBR.  As the rotational states are typically resolved (e.g. when probed by optical cycling), all significantly populated levels will need to be detected for accurate counting.

\begin{figure*}
    \centering
    \includegraphics[width=\textwidth]{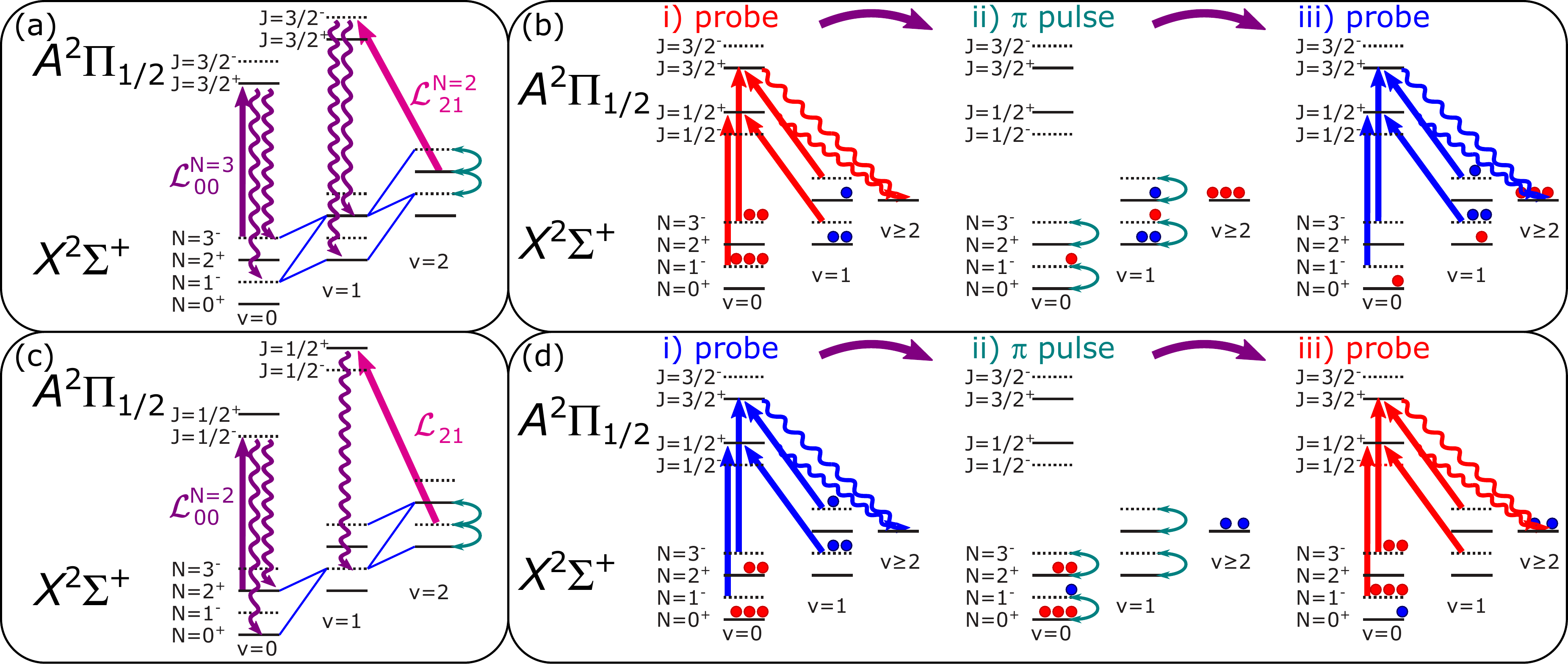}
    \caption{Two level thermalization and population measurement scheme for typical laser coolable molecule level structure. (a) Thermalization scheme for molecules initiated in $\ket{X^2\Sigma^+; v=0, N^P=1^-}$.  Solid (dashed) black lines denote $P=+$ ($P=-$) states.  Solid blue lines denote BBR-induced and spontaneous vibrational transitions,  straight arrows denote repump lasers, teal double arrows denote microwave transitions, and wavy arrows denote spontaneous optical decay paths.  (b) Method to independently measure the thermalized $v=0$ (red circles) and $v=1$ (blue circles) populations after thermalization scheme (a).  i) Four lasers address allowed optical cycling of the thermalized $v=0$ population, allowing population readout by laser induced fluorescence.  ii) Microwave $\pi$ pulses are applied to swap the parity $P$ of the thermalized populations.  iii)  The same four lasers used in step i) now optically cycle and detected the thermalized $v=1$ population.  (c) As in (a), but molecules initialized in $\ket{X^2\Sigma^+; v=0, N^P=0^+}$. (d) As in (b), with $v=1$ population probed before $v=0$.}
    \label{fig:two level molecule}
\end{figure*}

In laser-coolable molecules \cite{DiRosa2004}, 
this issue can be substantially mitigated by virtue of their nearly-diagonal Franck-Condon factors.   In Fig.\,\ref{fig:two level molecule}a we present a scheme which applies only two lasers and two microwave fields to close the system such that molecules initially in $\ket{X^2\Sigma^+;v=0,N^P=1^-}$ will evolve to only occupy $v=0$ and $v=1$. Laser $\mathcal{L}_{00}^{N=3}$ drives the $\ket{X^2\Sigma^+;v=0,N^P=3^-}\rightarrow\ket{A^2\Pi_{1/2}; v^\prime=0,J^{\prime P}=3/2^+}$ transition, which pumps population into $\ket{X^2\Sigma^+;v=0,N^P=1^-}$ to prevent excitation to states with $N>3$.  
Microwaves resonant with the transitions between $N=1,2$, and $3$ in the $v=2$ manifold allow transfer to the otherwise unoccupied $\ket{X^2\Sigma^+;v=2,N^P=2^+}$ state;
laser $\mathcal{L}_{21}^{N=2}$ then quickly depletes this state by driving the $\ket{X^2\Sigma^+;v=2,N^P=2^+}\rightarrow\ket{A^2\Pi_{1/2}; v^\prime=1,J^{\prime P}=3/2^-}$ transition. The combination of microwaves and laser $\mathcal{L}_{21}^{N=2}$ prevents significant population of states $v\ge2$. The two-photon microwave plus optical repumping out of $v=2$  ensures molecules will only spontaneously decay to $P=+$ parity $v=1$ levels populated by BBR-induced transitions. All laser and microwave fields are rapidly polarization modulated to destabilize dark $m_F$ levels.  

The equilibrium population distribution of this scheme has  only 2 significantly populated rotational states in each of $v=0,1$. Moreover, population transfer between $v=0,1$ only occurs due to BBR and spontaneous vibrational decay, and does not occur due to the repumping scheme (up to off-diagonal Frank-Condon factors).  The populations of excited states, as well as ground states with $v\ge2$, are roughly $(\Gamma_{01}+\Omega_{01})/\Omega_{\rm{R}}$, where $\Omega_{\rm{R}}$ is the Rabi frequency of the laser coupling to the state. In the limit of sufficient laser and microwave power to saturate the transitions, $\Omega_{\rm{R}}$ is typically $10^6$\,s$^{-1}$ to $10^7$\,s$^{-1}$. and the fractional population outside the effective two level system is less than $10^{-5}$ and can typically be calculated to better than 10\,\% uncertainty. Thus, the population rapidly approaches a thermal distribution for the $v=0,1$ two-level system at a temperature determined by the surrounding BBR.  Figure \ref{fig:two level molecule}c presents an equivalent two level thermalization scheme for molecules initiated in $\ket{X^2\Sigma^+; v=0, N^P=0^+}$.

Because blackbody radiation is incoherent, it is appropriate to model its interaction with quantum systems with a rate equation model using rates $\Gamma_{ij}$ and $\Omega_{ij}$.  We approximate the $v=0,1$ system as a two-level system with energy separation $\hbar \omega_{01}$, state populations $N_0$ and $N_1$, and total population $N=N_0+N_1$.  
Next, we abbreviate $\Omega_{01}=\Omega$ and $\Gamma_{01}=\Gamma$.  
The state populations evolve according to the rate equations
\begin{equation}
   \frac{d}{dt} \onebytwo{N_0}{N_1} = \twobytwo{-\Omega}{\Gamma+\Omega}{\Omega}{-\Gamma-\Omega} \onebytwo{N_0}{N_1}.
\end{equation}
If $N_0(t=0)=N$ and $N_1(t=0)=0$, then the solution is 
\begin{equation}\label{eq:N0}
    N_0 (t) = N \left( \frac{\Gamma+\Omega+\Omega e^{-t(\Gamma+2\Omega)}}{\Gamma+2\Omega} \right) ,
\end{equation}
\begin{equation}\label{eq:N1}
    N_1 (t) = N \left( \frac{\Omega(1- e^{-t(\Gamma+2\Omega)})}{\Gamma+2\Omega} \right) .
\end{equation}
The population evolves toward thermodynamic equilibrium with exponential time constant  $t_{\rm{therm}}= 1/(\Gamma+2\Omega)$ (in nuclear magnetic resonance notation, $t_{\rm{therm}}=T_1$).   Using the relation $\Omega/\Gamma = \left< n (\omega_{01})\right>$, we find that the asymptotic behavior is exactly that predicted for a thermal two-level system from quantum statistical mechanics:
\begin{equation}
    \frac{N_1}{N_0}= \frac{\left< n (\omega_{01})\right>}{\left< n (\omega_{01})\right>+1}.
\end{equation}

Importantly, the equilibrium state can be calculated from only  the temperature $T$ and energy separation $\hbar \omega_{01}$ of the states. Therefore, a population measurement of the two-level system at times which are long compared to $t_{\rm{therm}}$ is a quantum realization of the SI kelvin unit which is traceable to the second.  The equilibrium state is independent of the initial state distribution, as well as the experimentally poorly-known transition dipole matrix elements. However, measurement of the population dynamics on timescales comparable to $t_{\rm{therm}}$ in a uniform radiation field (such as a blackbody cavity) would enable the relevant transition dipole matrix element to be determined as well.

For laser-coolable molecules, 
it should be possible to utilize optical cycling to detect laser-induced fluorescence for independent, shot-noise-limited readout of the $v=0,1$ populations. Typically, optical cycling in molecules primarily occurs by laser $\mathcal{L}_{v,v^\prime}=\mathcal{L}_{00}$ driving the $\ket{v=0}\rightarrow\ket{v^\prime=0}$ transition.  Molecules decay from $v^\prime=0$ to ground vibrational states $v$ with branching fractions $b_{v^\prime=0,v}$, and ideally diagonal branching fractions $b_{v,v}\approx 1$.  Including $v=0$, the ground vibrational states with the first $k$ largest branching fractions are repumped by additional lasers \cite{DiRosa2004}.  If $v=v_{\rm{dark}}$, the vibrational state with the $(k+1)^{\rm{th}}$ largest branching fraction $b_{0,v_{\rm{dark}}}$, is not optically coupled, then roughly $N_\gamma=1/b_{0,v_{\rm{dark}}}$ photons may be scattered before the molecule reaches an uncoupled ``dark'' state.  For shot-noise-limited counting, the total photon collection efficiency $\eta$ should be $\eta\gtrsim b_{0,v_{\rm{dark}}}$.

In Fig.\,\ref{fig:two level molecule}b,d, we depict an optical cycling scheme coupling $v=0,1$ to readout the state populations after performing the thermalization schemes  in Fig.\ref{fig:two level molecule}a,c, respectively.  Using SrF as an example, the largest branching fraction to an uncoupled level would be $b_{02} \approx 4\times 10^{-4}$, allowing roughly $N_\gamma \approx 2500$ photons to be scattered before optically pumping into a dark vibrational state \cite{Barry2012}.  
The scheme can be generalized to  
using more repump lasers 
%
%
if necessitated by the branching fractions and detector efficiency.

 The standard molecule laser cooling scheme is closed to rotational branching by driving transitions between the first rotationally excited level of the ground electronic  state and the lowest rotational level of the excited electronic state \cite{Stuhl2008} (for transitions between two Hund's case a states, two Hund's case b states, or a ground Hund's case b to excited Hund's case a, the quantum numbers associated with such transitions are $J=1\rightarrow J^\prime=0$, $N=1\rightarrow N^\prime=0$, or $N=1 \rightarrow J^\prime=1/2$, respectively).  A feature of this optical cycling scheme is that all ground states, including vibrationally excited states, have $P=-$.  In contrast, BBR-induced excitation populates successively higher vibrational states with alternating parity (e.g.\ $P=(-1)^{v+1}$, assuming initial state $\ket{v=0,N^P=1^-}$).  Because vibrational excitation from optical cycling and BBR-induced transitions populate states with opposite parity, it is possible to probe the thermalized $v=0,1$ populations independently.

 In Fig.\,\ref{fig:two level molecule}b step i), optical cycling lasers are applied to first probe the thermalized $v=0$ population.  If cycling lasers are applied for a sufficient duration, all $v=0$ molecules scatter roughly $N_\gamma$ photons and are optically pumped to dark vibrational levels.  In step  ii), microwave $\pi$ pulses are applied to swap the parity $P$ of the thermalized populations in $v=0,1$ \cite{Williams2018,McCarron2018b}.  In step iii),  the optical cycling lasers are reapplied to detect the thermalized $v=1$ population.  Figure \ref{fig:two level molecule}d shows an equivalent population measurement scheme for molecules which thermalize under the scheme of Fig.\,\ref{fig:two level molecule}c.
 
 While it was assumed above that the optical cycling lasers are applied for a sufficient period to pump all addressed molecules to a dark state, this need not be the case.  If any optically addressed $P=-$ levels remain populated after step i), the $\pi$ pulse of step ii) will transfer the molecule to  optically dark $P=+$ levels.  
 
 For molecules with hyperfine structure, it may not be possible to  perform microwave $\pi$ pulses on every populated hyperfine manifold simultaneously.  In this case, steps ii) and iii) may be performed multiple times, with $\pi$ pulses performed on subsets of all hyperfine manifolds at each iteration.  Alternately, cycling lasers and microwave could be applied simultaneously to perform steps ii) and iii) at the same time, but might introduce systematic errors in population readout from non-identical cycling schemes  for the $v=0$ and $v=1$ populations.  Finally, we note that in addition to the BBR temperature measurment outlined below, the thermalization scheme depicted in Fig.\,\ref{fig:two level molecule}a,c could be applied to extend the trap lifetime $\tau_{\rm{trap}}$ of conservative molecule traps to exceed $1/\Omega$ \cite{McCarron2018b,Williams2018,Cheuk2018}.

We now assume the $\ket{0}$ and $\ket{1}$ state populations $N_0$ and $N_1$ can be measured independently, and construct a population asymmetry 
\begin{equation}\label{eq:asymmetry}
    \mathcal{A}(t)= \frac{N_0-N_1}{N_0+N_1}.
\end{equation}
For a shot noise-limited measurement, $\sigma_{N_0} = \sqrt{N_0}$, $\sigma_{N_1} = \sqrt{N_1}$, and 
\begin{equation}
\sigma_\mathcal{A}^2 =\frac{4N_0 N_1}{N^3}.
\end{equation}
Using Eq.\,\eqref{eq:stimulated rate} and Eq.\,\eqref{eq:asymmetry}, we can calculate the temperature uncertainty $\sigma_T$ 
\begin{equation}\label{eq:two level sensitivity}
    \sigma_T^2 = \left(\frac{\partial T}{\partial \Omega}\right)^2 \left(\frac{\partial \Omega}{\partial \mathcal{A}}\right)^2\sigma_ \mathcal{A}^2.
\end{equation}
Combining this with Eqs. (\ref{eq:spontaneous rate}), (\ref{eq:stimulated rate}), (\ref{eq:N0}), and (\ref{eq:N1}), 
and scaling the energy splitting to thermal energy by $x~=~\hbar\omega_{01}/k_{\rm{B}}T$, it is found that the temperature sensitivity of all two-level systems fall on a universal curve (Fig.\,\ref{fig:state_transfer_sensitivity}a solid lines) defined only by their transition energy $\hbar \omega_{01}$:
\begin{equation}\label{eq:variance T universal}
    \frac{\sigma_T}{T} = \sqrt{\frac{(e^x+1)^2}{e^{x}}}\frac{1}{x}\frac{1}{\sqrt{N}}.
\end{equation}

Equation (\ref{eq:variance T universal}) has a minimum fractional temperature uncertainty \mbox{$\sigma_T/T \approx 1.509/\sqrt{N}$} for $x\approx2.399$, i.e. optimal sensitivity occurs at temperature $T^*\approx\hbar\omega_{01}/2.399 k_{\rm{B}}$.  Moreover, the minimum of Eq.\,\eqref{eq:variance T universal} is fairly shallow, with \mbox{$\sigma_T/T$} reaching twice its minimum value at $x \approx 0.704$ and $x\approx 5.674$. Therefore, a single two level system has high sensitivity over a broad range of temperatures by the state transfer method.

\begin{table*}\centering
\begin{tabular}{rccccccc}
\hline\hline
         &$\nu_{01}$ (cm$^{-1}$) & $\mu_{01}$ (D)&$\Gamma_{01}$ (s$^{-1}$)    &$\Omega_{01}(T=300\,\rm{K})$ (s$^{-1}$)   &$t_{\rm{therm}}^{T=300\,\rm{K}}$ (s)  &$T^*$\,(K) \\ \hline
BaF                &468.9 \cite{NISTChemistry}                & 0.395  \cite{Vanhaecke2007a}                      &5.05                        &0.594                        &0.160  &281\\
AlCl                &481.3 \cite{NISTChemistry}                & 0.313 \cite{Yousefi2018}                        &3.42                        &0.378                        &0.239 &289 \\         

YbF              &501.91 \cite{NISTChemistry}                & 0.258 \cite{Vanhaecke2007a}                       &2.64                        &0.261                        &0.316  &301\\

SrF                &502.4 \cite{NISTChemistry}                & 0.264 \cite{Langhoff1986}                       &2.77                        &0.272                        &0.302  &301\\

CaF              &581.1  \cite{NISTChemistry}                & 0.275  \cite{HOU2018}                      &4.65                   &0.304              &0.190 &349\\

MgF               &711.69 \cite{NISTChemistry}                & 0.186   \cite{Hou2017}                     &3.91                   &0.133              &0.240  &427\\

AlF                &802.26 \cite{NISTChemistry}                & 0.235 \cite{Yousefi2018}                   &8.94                        &0.195                        &0.107  &481\\

YO          & 860.0 \cite{NISTChemistry} &  0.297 \cite{Smirnov2019} & 17.7  &0.289  &  0.0548 &516\\

BeH      &2006.1 \cite{Yadin2012}                &  0.094            &22.2 \cite{Yadin2012}                             &0.0015           &0.045 &1236\\
BH & 2366.9 \cite{NISTChemistry} & 0.088 \cite{BLINT1974} & 32.2 & 0.0004 & 0.031 & 1420\\

\hline\hline
\end{tabular}

\caption{Laser-coolable molecules with parameters relevant to BBR thermometry by state transfer.}
\label{tab:molecule data}
\end{table*}

In principle, any two-level system may use the state-transfer method outlined here to measure the temperature of the surrounding blackbody radiation.   Ideally, given some prior knowledge of the range of temperatures likely to be measured, a system should be chosen according to three criteria.  First,  $\omega_{01}$ should nearly minimize Eq.\,(\ref{eq:variance T universal}).  Second, $\mu_{01}$ should be sufficiently large that $t_{\rm{therm}}$ is short compared to the timescale of a measurement. Third and finally, technical considerations should be given to maximize the number of identical measurements $N$.  

In Table \ref{tab:molecule data}, we present the relevant vibrational transition parameters to determine $t_{\rm{therm}}$ and $\sigma_T/T$ for several laser coolable molecules of interest.  Transition dipole matrix elements $\mu_{01}$ are as given in the associated reference, or else calculated using
\begin{equation}
	\label{eq:vib_matrix_element}
	\mu_{v,v+1} = \left(\frac{v+1}{2}\right)^{1/2}\left(\frac{\hbar}{m\omega_e}\right)^{1/2}\left[\frac{d\mu}{dR}\right]_{R=R_e},
\end{equation}
where $\omega_e$ is the vibrational frequency, $[d\mu/dR]_{R=R_e}$ is the derivative of the dipole moment with bond length $R$, and $m$ is the reduced mass.

Assuming typical  $N\approx 10^4$ molecules at a repetition rate of $\mathcal{R}=10$\,s$^{-1}$, temperature sensitivity  $\sigma_T/T \approx 10^{-4}$ could be achieved in a $t\approx 1$\,h, comparable to the time to equilibrate a blackbody cavity with a fixed-point reference.  New techniques, such as loading a MOT from a continuous beam \cite{shaw2020} or molecule Zeeman slower \cite{Petzold2018} could enable significantly more trapped molecules to push the measurement sensitivity to $\sigma_T/T \approx 10^{-5}$ or lower.

\subsection{Frequency Shifts}
To first order, the rovibrational energy of a diatomic molecule is given by
$\hbar \omega_{v,N} = BN(N+1) +\hbar \omega_e(v+1)$, where $B$ is the rotational constant, $\omega_e$ is 
the 
vibrational constant, $N$ is the rotational quantum number, and $v$ is the vibrational quantum number.    In order to understand the relative contributions of the rotational and vibrational structure to the BBR frequency shift, we first consider only the rotational structure, estimating the shifts on rotational state $N$ from all other rotational states.  The transition energies between neighboring rotational levels are $\hbar\omega_{N,N-1} = 2BN$.  Typical rotational constant values are $B\sim h\times 10$\,GHz, while $k_{\rm{B}} T \approx h \times 6$~THz for $T=300$\,K.  Thus,  for reasonably small $N$, $x_{N,N^\prime} = \hbar \omega_{N,N^\prime}/ k_{\rm{B}}T \ll 1$ near room temperature, and we can use the approximation of \eqref{eq:farley_wing_approx}.   In this limit, the shift is given by
\begin{equation}
    \label{eq:rotational_shifts}
    \Delta \omega_{N} \approx -\frac{B \mu^2}{9\epsilon_0\hbar^2 c^3}\left(\frac{k_B T}{\hbar}\right)^2.
\end{equation}
Note that in this model, the BBR shift is independent of rotational state $N$.  Therefore, the shifts in neighboring rotational levels cancel (up to non-rigid rotor terms $\sim~DN^2(N+1)^2$).  Measurement of the differential BBR shift between neighboring rotational levels is therefore  unlikely to be a generally viable thermometry scheme.

The vibrational transition frequencies $\omega_{v,v^\prime}$ in a molecule are generally  much closer to the peak of the blackbody spectrum at room temperatures.   Inserting \eqref{eq:vib_matrix_element} into \eqref{eq:full_freq_shift} and considering coupling between $v$ and $v^\prime = v\pm 1$, we find
\begin{equation}
    \label{eq:vib_shift}
	\Delta \omega_v=\frac{1}{6\pi^2 \epsilon_0 \hbar c^3}\left(\frac{\hbar}{m\omega_e}\right)\left[\frac{d\mu}{dR}\right]^2_{R=R_e}\left(\frac{k_B T}{\hbar}\right)^3\mathcal{F}\left(\frac{\hbar\omega_e}{k_B T}\right).
\end{equation}
Note that for $v=0$, this expression is also correct even though the only nonzero matrix element is $\mu_{01}$.  Much like the rigid rotor example, the harmonic oscillator experiences a state-independent frequency shift up to anharmonic terms.


Nonetheless, we estimate the sensitivity for BeH and MgF from calculated potential energy curves, which implicitly include all anharmonic terms.   Eigenenergies and dipole matrix elements $\mu_{ij}$ are taken from the ExoMol line lists for for MgF \cite{Hou2017} and BeH \cite{Yadin2012}.  For measurements limited by  coherence time $\tau$, the fractional sensitivity to $T$ of a frequency shift measurement is given by
\begin{equation}\label{eq:shift sensitivity}
	\frac{\sigma_T}{T} = \frac{1}{T}\left(\frac{\partial T}{\partial \Delta \omega_{v,N,v^\prime,N^\prime}^\prime}\right) \frac{1}{\tau},
\end{equation}
where $\Delta \omega_{v,N,v^\prime,N^\prime}^\prime=\Delta\omega_{v^\prime,N^\prime}-\Delta\omega_{v,N}$ is the differential frequency shift when considering the full rovibrational structure.  In Fig. 1b, we evaluate \eqref{eq:shift sensitivity} for the $\ket{v=0,N=1}\rightarrow \ket{v^\prime=1,N^\prime=0}$ transition.   The coherence time is assumed to be limited by the state lifetime $\tau=1/\gamma_{01} = 1/(\Gamma_{01}+\Omega_{01})$, with  $\Gamma_{01}$ listed in Table \ref{tab:molecule data} and $\Omega_{01} = \Gamma_{01}\langle n(\nu_{01})\rangle$.

As shown in Fig.\,\ref{fig:state_transfer_sensitivity}, the fundamental temperature sensitivity of frequency shift measurements at the coherence time limit may exceed that of the state transfer method at high temperatures.  However, it should be noted that achieving lifetime-limited linewidths in molecules at present would be a daunting task.
For example,  in MgF at $T\approx 600$\,K, interrogation for time $\tau\approx0.2$\,s achieves fractional sensitivity $\sigma_T/T\approx 1/\sqrt{N}$.  This implies a quality factor $Q = \omega/\gamma \approx 10^{13}$, about an order of magnitude higher than has been achieved in a molecular lattice clock \cite{Kondov2019}.  Reaching $10^{-5}$ fractional temperature uncertainty would require controlling systematic frequency shifts at the $10^{-18}$ level, comparable to the best atomic frequency standards.  Indeed, existing atomic frequency standards with $10^{-18}$ fractional frequency uncertainty now require thermodynamic temperature measurements with $2\times10^{-5}$ accuracy \cite{Beloy2014}.


\section{Rydberg Atoms}
Rydberg atoms offer interesting trade-offs compared to molecules for sensing BBR.  Rydberg atoms have transition dipole moments which are three or more orders of magnitude larger than molecules (with interaction strength proportional to the square of the dipole moment).  On the other hand, molecules are sensitive to electromagnetic fields in their ground electronic state and thus may gain sensitivity through many orders of magnitude longer interaction time.   For Rydberg atoms, the transition dipole matrix elements $\mu_{ij}$ are determined by  quantum defect theory at the $\sim$~1~\% level, while there is little experimental data for vibrational transition dipole matrix elements in molecules.  Therefore,  BBR temperature may be determined by dynamic, as well as equilibrium, population measurements in Rydberg systems.

Rydberg atoms were first used to measure BBR temperature by Hollberg and Hall in 1984 with a roughly 20\,\% absolute uncertainty \cite{Hollberg1984}.
BBR temperature can be determined in Rydberg systems by measuring bound-to-bound \cite{Figger1980, Beterov2009} or bound-to-continuum \cite{Spencer1982} transition rates.  Ovsiannikov \textit{et al.} have considered using Rydberg atoms to sense the BBR temperature for more accurate corrections to optical lattice clocks \cite{Ovsiannikov2011}. 

Significant progress has been made in the development of  Rydberg-atom spectroscopic approaches for radio-frequency electric field strength measurements 
\cite{holl1, holl2, sed1, fan1, simons1, simons2, holl3, gordon1}. 
Rydberg atoms can be used to detect THz radiation, such as thermal radiation. Electromagnetically induced transparency (EIT) is a two-photon process utilized for efficient Rydberg excitation by coupling a ground state to the Rydberg state via an intermediate state. This nonlinear process can also be utilized for THz field sensing. BBR-induced bound-to-bound transition rates may be measured via state-selective electric-field ionization and counting of the field electrons (or ions).  Thermal ionization rates may also be measured via photo-electron or ion counting.  

\begin{figure}
\centering
\scalebox{.4}{\includegraphics*{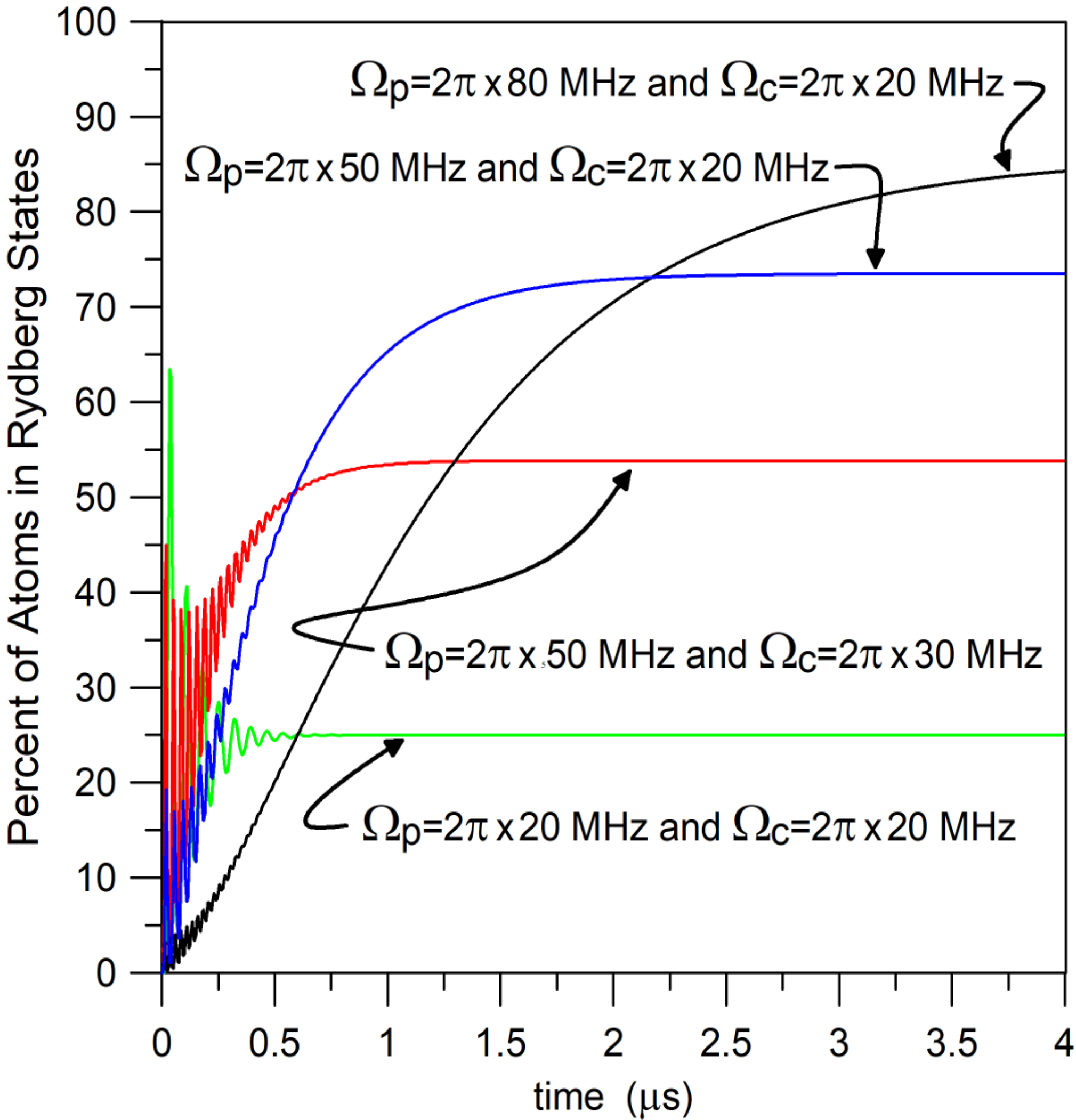}}\\
\caption{Rb Rydberg state population for a two-photon excitation.  Lines are labeled according to the Rabi frequencies $\Omega_{\rm{p}}$ and $\Omega_{\rm{c}}$ of the probe ($\ket{5S}\rightarrow\ket{5P}$) and coupling ($\ket{5P} \rightarrow \ket{nS}$) transition lasers, respectively. }
\label{pol}
\end{figure}

A two-photon excitation scheme has been demonstrated as  a simple and efficient method to produce Rb Rydberg atoms. In Rb for example,  excitation via $5S_{1/2} \rightarrow 5P_{3/2} \rightarrow nD_{3/2}$ (where $n>15$) may transfer  $60~\%$ to $90~\%$ of the resonant atoms to populate the Rydberg state, which is illustrated in Fig.~\ref{pol}. This figure shows the fraction of atoms excited to Rydberg states as a function of time for different $5S_{1/2}\rightarrow P_{3/2}$ laser powers.  These results are obtained from the solution of the master equation for the density matrix components of a three-level atomic system discussion in \cite{holl4}. 

There are, in principle, three BBR effects on Rydberg atoms that may be considered: level shifts, BBR-induced bound-to-bound  transitions, and BBR-induced photo-ionization. The BBR shift of Rydberg levels is only on the order of 10 Hz/K near room temperature. It also tends to be the same for all high-lying Rydberg states, because much of the shift is due to the ponderomotive energy shift of the quasi-free Rydberg electron in the thermal radiation field. In principle one may consider optical measurements of the shifts of individual Rydberg levels, using an auxiliary clock-stabilized Rydberg excitation laser, or a microwave measurement of the transition frequencies between pairs of electric-dipole-coupled Rydberg levels. Microwave measurements of transitions between Rydberg levels would have to be performed at quite low principal quantum numbers ($n \sim 15$), where the thermal Rydberg level shifts become state dependent. In any case, the task would involve measuring transition shifts in the range of \mbox{10\,mHz} to \mbox{10\,Hz}, for Rydberg excitation lines that are \mbox{1\,kHz} or more wide. Assuming that this can be done, one would obtain a calibration-free, atom-based measure of the BBR. 

Due to technical and signal-to-noise challenges associated with the level-shift approach, promising alternatives for BBR sensing are to measure BBR-induced Rydberg photoionization rates and Rydberg bound-to-bound transition rates. Thermal ionization rates are in the range of $10^3$\,s$^{-1}$, while bound-to-bound transitions can have rates between $10^4$\,s$^{-1}$ to $10^5$\,s$^{-1}$ , for each final state. Typically, there are on the order of ten final states that become significantly populated. Operating on sample sizes of $10^6$ to $10^{10}$ Rydberg atoms, summed over $10^4$ to $10^6$ experimental cycles acquired at a data rate of about 100\,s$^{-1}$, we anticipate sufficient statistics to measure temperature to $\sigma_T/T \lesssim 10^{-5}$.


\subsection{State Transfer}
\subsubsection{Bound-to-bound state transfer}\label{ss:bb st}
Here we consider measurements involving BBR-induced bound-to-bound state transfer in Rydberg atoms.  In these measurements, we excite atoms to an $nS$ Rydberg state using a pulse of the order of 2\,$\mu$s, wait a variable time for atoms to evolve in the BBR field, then induce a field-ionization pulse to analyze the transfer of atoms to surrounding $P$ states.  Counting the distribution of resulting ions then constitutes a measurement.  A benefit of Rydberg atoms is that large numbers of Rydberg atoms ($N\sim 10^7$) may be prepared with high repetition rate ($\mathcal{R}\sim 100$~s$^{-1}$). For example,  a vapor cell with at 60 $^\circ$C has a Rb density $\approx 3\times 10^{11}$\,cm$^{-3}$ (the cell may be thermally isolated from the BBR and need not be at the temperature $T$ of the surrounding blackbody).  We have confirmed our estimates of $N$ by using a time-domain analysis of the four-level system \cite{holl4}. We have used this same scheme, with great success, to generate Rydberg atoms for  atom-based electric field sensing \cite{holl1, holl2, sed1, fan1, simons1, simons2, holl3, gordon1}.

We model this interaction by assuming $N$ atoms are initiated in an excited  Rydberg state $\ket{0}$.  This state has partial spontaneous decay rates $\Gamma_{0j}$, a BBR-stimulated transfer rate $\Omega_{0j}$ from $\ket{0}$ to other, nearby states (generally labeled $\ket{j}$) that we can measure. For concreteness, we will consider the case of $N$ Rb Rydberg atoms in the  $ nS \equiv \ket{0}$ state at $t=0$.  Decay from $nS$  significantly populates several states, with $nP$ and $(n-1)P$ achieving the largest peak populations \cite{Galvez1995}.  We choose to probe the $(n-1)P\equiv\ket{1}$ population dynamics, as it is typically eaisier to resolve from $nS$ than $nP$ through selective ionization. 
As atoms leave the initially populated $nS$ level, multi-step processes or ``cascading'' can result in relatively large populations in additional states.  Cascading is  known to affect the $(n-1)P$ state population dynamics at the 1~\% level, and is ignored in the analysis here.
  
   Defining the total depopulation rate of state $\ket{i}$ to be
   \begin{equation}
      \gamma_i = \sum_j\left(\Gamma_{ij}+\Omega_{ij}\right),
   \end{equation}
   the population of state $\ket{1}=(n-1)P$ is given by Ref.~\cite{Galvez1995}, Eq. (5):
\begin{equation} \label{eq:Rydberg N1}
    N_1 = N\frac{\Omega_{01}+\Gamma_{01}}{\gamma_1-\gamma_0}\left(e^{-\gamma_0 t} -e^{-\gamma_1 t}\right).
\end{equation}
The population $N_1$ is maximized at time
\begin{equation}
    t^\text{max} = \frac{1}{\gamma_1-\gamma_0}\ln{\frac{\gamma_1}{\gamma_0}}.
\end{equation}
Evaluating $N_1(t=t^{\rm{max}})$ yields 
\begin{equation}
    N_1^\text{max} = N\frac{\Omega_{01}+\Gamma_{01}}{\gamma_0}\left(\frac{\gamma_0}{\gamma_1}\right)^{\frac{\gamma_1}{\gamma_1-\gamma_0}}.
\end{equation}

\begin{figure}
    \centering
    \includegraphics[width=\columnwidth]{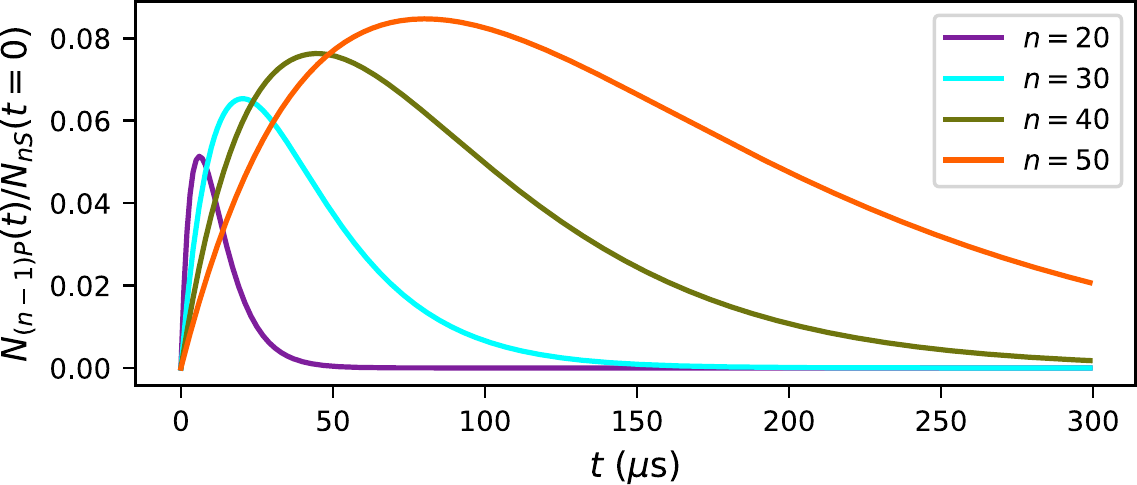}
    \caption{\label{fig:Rydberg pop vs timel} Rydberg state population transfer dynamics under the three level model.}
\end{figure}

Figure\,\ref{fig:Rydberg pop vs timel} shows $N_1$ as a function of time calculated using Eq.\,\eqref{eq:Rydberg N1}, with dipole matrix elements and transition energies calculated using the Alkali Rydberg Calculator python package \cite{ARC2020, NISTDisclaimer}. Unlike the molecule case of Sec III, Rydberg atoms are not closed two level systems, making the timing of the state measurement important. Operationally, given an expected typical temperature $T^*$, the population $N_1$ should be measured at a various times around the expected maximizing time $t^{\rm{max}*}$.  These measurements may then be fit to Eq.~\eqref{eq:Rydberg N1} in order to determine $\Omega_{01}$, and thus estimate $T$.

We may estimate the sensitivity of this procedure by first noting that it is typically the case in Rydberg systems that $x_{01}= \hbar \omega_{01}/k_{\rm{B}} T \ll 1$.  Therefore, using Eq.\,(\ref{eq:stimulated rate}),
\begin{equation}
\Omega_{01}\approx \frac{\mu_{01}^2\omega_{01}^2}{3\varepsilon_0\pi\hbar^2 c^3}k_{\rm{B}}T \equiv a T,
\end{equation}
which is linear in temperature.

If we measure the population $N_1$ at 
time $\tau \ll t^{\rm{max}}$, then
\begin{equation}
\begin{aligned}\label{eq:Rydberg N1 approx}
	N_{1} &\approx N \Omega_{01} \tau\\
	&= N a \tau T.
\end{aligned}\end{equation}
We then find
\begin{equation}\label{eq:Rydberg state trans sensitivity}
	\frac{\sigma_T}{ T} \approx \sqrt{\frac{1}{N_1}+\frac{1}{N}}.
\end{equation}
While Eq.~\eqref{eq:Rydberg N1 approx} is not strictly valid at $\tau = t^{\rm{max}}$, we can estimate the optimal sensitivity of the Rydberg state transfer method by substituting $N_1=N_1^{\rm{max}}$ into Eq~\
\eqref{eq:Rydberg state trans sensitivity}.  The calculation is shown as dashed lines in Fig.\,\ref{fig:state_transfer_sensitivity} for Rb $nS$ states, with $n=20$ to $n=50$.  The fractional temperature sensitivity of this method in the shot noise limit is comparable to the molecule state transfer measurement, with $\sqrt{N}\sigma_T/T \approx 4 - 5$ for a wide range of temperatures $T=100$\,K to $T=1300$\,K.

Interrogation by two counter-propagating laser beams over a cell length of 2 cm with a $1/e^2$ beam diameter 0.5 cm should excite $4\times 10^7$ atoms into a Rydberg $nS$ state in a velocity class near the center of the 230 MHz Doppler profile.   Of the atoms that are excited, roughly 6\,\% or $2\times 10^6$ atoms will be transferred to an adjacent $P$ state after one state lifetime.    Assuming both shot-noise-limited detection and repeating the experiment at $\mathcal{R}=100\,$s$^{-1}$, BBR temperatures $T\,\approx \, 300$\,K may be determined with $\sigma_T\approx 100$\,mK uncertainty for 1\,s of averaging.

Achieving shot noise-limited readout poses a large challenge, given that the ion current is only about \mbox{300~fA} per pulse.  This should be compared with the current require to charge two plates to the necessary electric field strength of 1 kV/cm for inducing the field ionization:  for 2 cm $\times$ 1 cm plates separated by 3.5 mm, the capacitance is approximately 0.5\,pF.  To charge 0.5\,pF to 500\,V in 10\,$\mu$s requires 20\,$\mu$A.  Thus, the vapor cell needs to control stray electric fields and minimize leakage currents to prevent any contamination of the ion signal.  A specialized ammeter capable of separating such large and small currents signals was discussed in Ref.\,\cite{Eckel2012}.

\subsubsection{Ionization}
In this section, we consider BBR-induced bound-to-continuum Rydberg atom transitions, where $N$ atoms  initially in state $\ket{0}=nL$ are ionized (i.e.\ transferred to state $\ket{1}$) at rate $\mathcal{W}$, and  decay to all other states $\ket{j}$ at rate $\gamma_{01}$.
We use the approximate Eq.~(27) derived in Ref.~\cite{Beterov2009}, to calculate the ionization rate of an alkali atom,
\begin{equation}\label{eq:ionization rate}
\begin{aligned}
\mathcal{W}& = \frac{B_L T}{(n-\mu_L)^{7/3}}\ln{\frac{1}{1-\exp(-\frac{R_\infty}{k_{\rm{B}}T(n-\mu_L)^2})}}\\
&\times \Big( \cos(\Delta_L^+ +\frac{\pi}{6})^2 +\cos(\Delta_L^- -\frac{\pi}{6})^2 \Big),  
\end{aligned}
\end{equation}
where  $B_L=11500 A_L$\,K$^{-1}$\,s$^{-1}$,  $A_L$ is a scaling coefficient of order unity, 
$\mu_L$ is the quantum defect, 
$\Delta_L^+ = \pi(\mu_L - \mu_{L+1})$, and $\Delta_L^- = \pi(\mu_{L-1}-\mu_L)$.  Values for these parameters are taken from Refs.~\cite{Beterov2009,Li2003}.  For Rb $nS$ states, we take $A_L=1$, $\Delta_L^+ = \pi\times0.490134$, $\Delta_L^- = 0$, and $\mu_L = \mu_0+\mu_2/(n-\mu_0)^2$, where $\mu_0=3.1311804$ and $\mu_2=0.1784$.

 The state populations are governed by the rate equations 
\begin{equation}\label{eq:ion rate equations}
\begin{aligned}
\frac{dN_0}{dt} & =  -(\gamma_{0j}+\mathcal{W}) N_{0},\\
	\frac{dN_1}{dt} & =  \mathcal{W} N_{0}, \\
	\frac{dN_j}{dt}&= \gamma_{0j} N_0.
\end{aligned}\end{equation}
As with the molecule state transfer thermometry method of Sec. IIIA, an advantage of the ionization thermometry method is the atomic population evolves toward an equilibrium distribution.  In the $t\rightarrow\infty$ limit,
\begin{equation}
	N_1 = N \frac{\mathcal{W}}{\mathcal{W}+\gamma_{0j}}.
\end{equation}

The total number of ions produced is typically $N_1/N \approx 0.1$, similar to $N_1^{\rm{max}}/N$ in the bound-to-bound method of Sec. \ref{ss:bb st}.  However, the photo-ionization method here requires only a modest guiding field to detect ions with high efficiency.  Therefore, achieving the shot noise limit should be less technically challenging than the bound-to-bound state transfer method, which requires $\sim 1$\,kV ionizing field for detection. We estimate the temperature sensitivity of BBR-induced ionization in the shot noise limit as

\begin{equation}\label{eq:ion sensitivity}
\begin{aligned}
    \sigma_T^2 &= \Big( \frac{\partial T}{\partial \mathcal{W}}\Big)^2 \left[ \Big(\frac{\partial \mathcal{W}}{\partial N_1}\Big)^2 {N_1}
                                                                              +\Big(\frac{\partial \mathcal{W}}{\partial N  }\Big)^2 {N  } \right] \\
                                                                      &= \Big(\frac{\partial T}{\partial \mathcal{W}}\Big)^2 \frac{\gamma_{0j}^2}{(1-N_1/N)^4}\frac{N_1 N +N_1^2}{N^2}\frac{1}{N}\\
               &= \Big(\frac{\partial T}{\partial \mathcal{W}}\Big)^2 \frac{ \mathcal{W}( \mathcal{W}+\gamma_{0j})^2(2 \mathcal{W}+\gamma_{0j})}{\gamma_{0j}^2}\frac{1}{N}.         
 \end{aligned}
\end{equation}                                                                           
Using Eq.~\eqref{eq:ion sensitivity}, calculated values of $\sqrt{N}\sigma_T/T$ for Rb $nS$ states by the ionization method are shown as dotted lines in Fig.~\ref{fig:state_transfer_sensitivity}.  The simple three level model outlined here is expected to slightly underestimate the Rydberg ionization temperature sensitivity, as we have ignored BBR induced ionization from other states $\ket{j}$.  Bound-to-bound transitions from $nS$  are primarily to states with neighboring principal quantum number $n$,  which ionize at a rate of similar order of magnitude to the $nS$ state.  By calculating $\mathcal{W}$ with a fuller modeling of the Rydberg population dynamics than Eqs.~\eqref{eq:ion rate equations}, we  expect the error due to the three level approximation in \eqref{eq:ion sensitivity} to be $<10$~\% at room temperature.
\subsection{Frequency Shifts}

\begin{figure}[t!]
    \centering
    \includegraphics[width=\columnwidth]{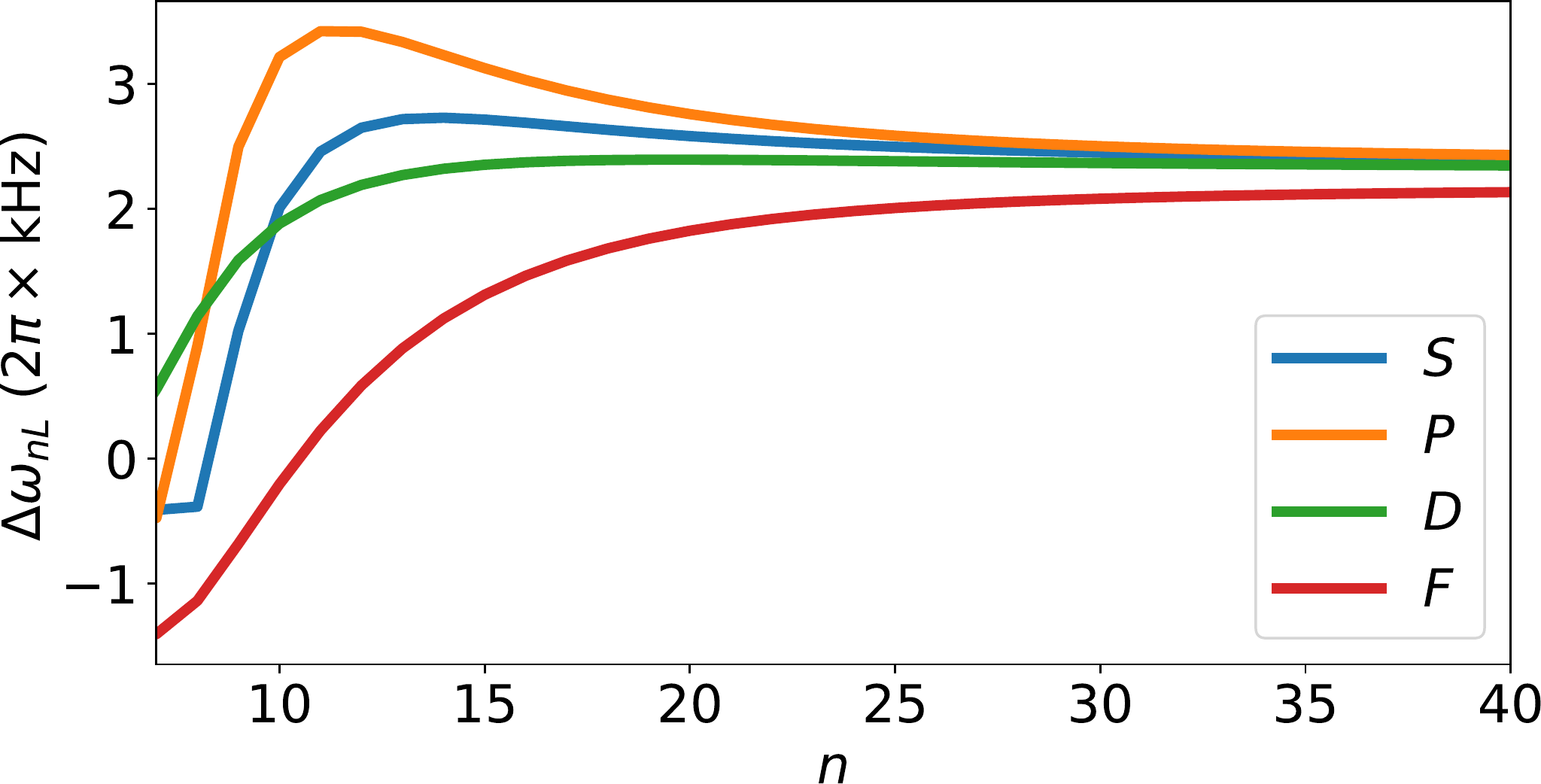}
    \caption{\label{fig:rydberg_shifts} BBR shift at $T\,=\,300$~K of Rb Rydberg states as a function of principle quantum number $n$, calculated using Eq.~\eqref{eq:rydberg_shift}.}
\end{figure}

We consider the energy $\hbar \omega_{nL}$ of state $\ket{nL}$, shifted by the BBR interaction with all other states $\ket{n'L'}$:
\begin{equation}
    \label{eq:rydberg_shift}
     \Delta \omega_{nL} = -\frac{1}{6\pi^2 \epsilon_0 \hbar c^3}\left(\frac{k_B T}{\hbar}\right)^3\sum_{n'L'} \mu_{nL,n'L'}^2 \mathcal{F}\left(\frac{\hbar \omega_{nL,n'L'}}{k_B T}\right),
\end{equation}
where $L'=L\pm1$ and $\omega_{nL,n'L'} = \omega_{n'L'} - \omega_{nL}$.  For large  $n$, $\hbar \omega_{nL,n'L'}/k_B T \ll 1$.  In this case, using the oscillator sum rule  and the approximation of Eq.\,\eqref{eq:farley_wing_approx} for $\mathcal{F}(y)$ yields
\begin{equation}
	\label{eq:bbr:rydberg_asymptotic_shift}
     \hbar \Delta \omega_{nL} \approx \frac{\pi}{3}\alpha^3\frac{\left(k_B T\right)^2}{E_h},
\end{equation}
where $\alpha\,\approx\,1/137$ is the fine structure constant.  The shift is proportional to $T^2$ and is about 2.4\,kHz at $T=300$\,K.  The shifts for some Rb Rydberg levels are evaluated using Eq.~\eqref{eq:rydberg_shift} and shown in Fig.~\ref{fig:rydberg_shifts}.  For Rb, the shift $\Delta \omega_{nL}$  at room temperature is roughly 1000 times larger in Rydberg states than the $5P$ states.  Therefore, the differential shift of the $5P\rightarrow nL$ is essentially equal to $\Delta \omega_{nL}$. The resulting fractional temperature sensitivities to for Rb $nS$ states are shown in Fig.~\ref{fig:state_transfer_sensitivity}b, assuming lifetime-limited interaction time.

While technically more challenging than state transfer, Rydberg frequency shift measurements offer superior sensitivity in the high temperature regime. 
Moreover, frequency shift thermometry in Rydberg systems appears favorable compared to molecular systems.
First, the shifts are larger by roughly three orders of magnitude (e.g.\ fractional temperature uncertainty of $10^{-5}$ requires only $10^{-15} - 10^{-16}$ fractional frequency uncertainty).
Having comparable sensitivity $\sqrt{N} \sigma_T/T$ to molecules but with relatively short state lifetimes, frequency shift measurements in Rydberg systems may be repeated more rapidly $\mathcal{R}\sim 1000$\,s$^{-1}$.
In this case, with $N=1000$ atoms per measurement, $10^{-5}$ fractional temperature uncertainty could be achieved in $t\sim$1\,h averaging time.
Frequency measurements of Sr and Yb Rydberg states in an optical lattice at similar sensitivity have been proposed a method for reducing the BBR shift uncertainty in atomic clocks~\cite{Ovsiannikov2011}.

\section{Conclusion}

The successful implementation of a Rydberg or molecule-based BBR detector would provide an entirely new and direct path to establish primary, quantum-SI-compatible measurements for both radiation and temperature.  Development of these techniques is a promising means to correcting BBR frequency shifts in optical clocks \cite{Ovsiannikov2011}.
Furthermore, integration of this technology with 
cold atom miniaturization \cite{Barker2018,McGilligan2020} 
programs 
will enable quantum SI radiometry and thermometry to be realized within a single laboratory, or even deployed in a mobile platform. Mobile standards could find several applications, such as on-board primary radiometry calibrations for remote-sensing satellites \cite{Gu2012,Datla2014} 
and high-accuracy non-contact thermometers.

\section*{Acknowledgement}
The authors thank Kyle Beloy, Dazhen Gu, Andrew Ludlow, Georg Raithel, Matt Simons, and Howard Yoon for insightful conversations,  and thank Alexey Gorshkov, \mbox{Nikunjkumar} Prajapati, and Wes Tew  for careful reading of the manuscript.  We are grateful to David DeMille, Shiqian Ding, Daniel McCarron, Micheal Tarbutt, and Jun Ye, who made us aware of several papers referenced in this work.  This work was supported by NIST.

\bibliography{thebib}

\begin{thebibliography}{68}
\bibitem [{\citenamefont {{Any mention of commercial products within this work
  is for information only; it does not imply recommendation or endorsement by
  NIST}}()}]{NISTDisclaimer}%
  \BibitemOpen
  \bibfield  {author} {\bibinfo {author} {\bibnamefont {{Any mention of
  commercial products within this work is for information only; it does not
  imply recommendation or endorsement by NIST}}},\ }\href@noop {}
  {}\BibitemShut {NoStop}%
\end{thebibliography}

\end{document}